\documentclass [12pt,preprint]{aastex}
\usepackage{epsfig}
\begin{document}
\voffset-1cm
\newcommand{\gsim}{\hbox{\rlap{$^>$}$_\sim$}}
\newcommand{\lsim}{\hbox{\rlap{$^<$}$_\sim$}}

\title{Gamma Ray Bursts, Supernovae and Metallicity\\
in the Intergalactic Medium}

\author{Shlomo Dado\altaffilmark{1}, Arnon Dar\altaffilmark{1} and A. De
R\'ujula\altaffilmark{2}}

\altaffiltext{1}{dado@phep3.technion.ac.il, arnon@physics.technion.ac.il,
dar@cern.ch.\\
Physics Department and Space Research Institute, Technion, Haifa 32000,
Israel.}
\altaffiltext{2}{alvaro.derujula@cern.ch; Theory Unit, CERN,
1211 Geneva 23, Switzerland. \\
Physics Department, Boston University, USA.}

\begin{abstract}
The mean iron abundance observed in the intergalactic medium (IGM) 
within galaxy clusters and without galaxy clusters is consistent with the mean 
amount of iron per unit volume in the Universe which has been produced by 
standard supernova (SN)  explosions with a rate proportional to the cosmic 
star-formation rate. If most SNe took place inside galaxies, then the IGM could 
have been enriched with their metals by galactic winds and jets that swept most 
of the galactic gas with the SNe ejecta into the IGM. A significant fraction of 
the early SNe, however, could have taken place outside galaxies or within dwarf 
galaxies, which were later disrupted by tidal interactions, and/or mass loss 
through fast winds, SN ejecta and jets.  Little is known about such 
intergalactic SNe at high red-shifts. They could have occurred primarily in 
highly obscured environments, avoiding detection. Supporting evidence for 
intergalactic SNe is provided by SNe associated with gamma ray bursts (GRBs) 
without a host galaxy and from the ratio of well localized GRBs with and without 
a host galaxy. A direct test of whether a significant contribution to the iron 
abundance in the IGM came from intergalactic SNe would require the measurement 
of their rate per comoving unit volume as function of red-shift.  This may be 
feasible with IR telescopes, such as the Spitzer Space Telescope.

\end{abstract}

\section{Introduction}

\noindent 
Iron abundances in the hot intracluster medium (ICM) in galaxy 
clusters and in the intergalactic medium (IGM) have been known to be 
rather enriched: their typical abundance is one third of the solar one 
(e.g.~Edge \& Stewart~1991; Arnaud et al.~1992). 
Recent precise measurements of intracluster abundances with Beppo-SAX, 
Chandra and XMM-Newton have confirmed that the average metallicity 
throughout most of the volume of galaxy clusters is $Z\sim 0.3\, Z_\odot$ 
(e.g. Balestra et al.~2007 and references therein). It has 
been argued that such a metallicity is several times larger than that 
expected from standard supernova (SN) explosions assuming standard initial 
mass functions (e.g., Renzini et al.~1993; Brighenti \& 
Mathewes~1998; Maoz \& 
Gal-Yam~2004). It was suggested (Scannapieco, Schneider \& 
Ferrara~2003 and references therein) that such a 
puzzling iron enrichment of the ICM may be due to the production and 
dispersion of metals from explosions of Population III stars 
--hypothetical, extremely massive and hot stars with virtually no metal 
content, believed to have existed in the early universe. These stars have 
been invoked to account for faint blue galaxies, and the heavy elements in 
quasar emission spectra, which cannot be primordial. It has also been 
suggested that Population III stars triggered the period of reionization 
(Loeb \& Barkana~2001 and references therein) as inferred from 
the measured polarization of the cosmic microwave background radiation 
(Barkats et al.~2005). So far, Population III stars have not 
been observed\footnote{Indirect evidence for the existence of Population 
III stars has been claimed from gravitational lensing of a high red-shift 
galaxy (Fosbury et al.~2003)}.
In this note we present a simple calculation which shows that the mean 
iron abundance observed in the IGM within galaxy 
clusters and without galaxy clusters is consistent with the mean amount of 
iron per unit volume in the Universe which has been produced by standard 
SN  explosions with a rate proportional to the cosmic 
star-formation rate (see also, e.g., Maoz \& Gal-Yam~2004). 
Such SNe could have produced the iron abundance in the ICM and IGM with no 
need for Population III stars, provided that their metals were transported 
and dispersed in the IGM.  Indeed, in the local universe, the total mass of 
gas in spiral and elliptical galaxies is much smaller than the total mass 
in stars, whereas in galaxy clusters, approximately 1/6 of the 
baryons reside in stars within galaxies, while 5/6 are 
in the ICM (e.g., Ettori~2003). The simplest interpretation 
of these observations is that most of the gas in galaxies was swept into the 
IGM by winds and jets produced by star formation and stellar evolution. 
in galaxies. These winds could have transported into the IGM the metals 
that were injected into the ISM of gallaxies by galactic SN explosions.

In the local universe, most SN explosions take place inside galaxies. 
However, a 
significant fraction of the early SNe could have taken place also outside
galaxies or within dwarf galaxies, which were later disrupted 
by tidal interactions, and/or mass loss through fast winds, SN ejecta and 
jets. Recent near-infrared (Perez-Gonzalez et al.~2005) and 
radio searches for SNe have shown that at higher redshifts SNe took place 
mostly in very dusty environments and that their rate per comoving unit 
volume increased sharply with red-shift $z$ (e.g., Mannucci et 
al.~2007). Such intergalactic 
SNe could have produced a significant fraction of the metals in the IGM. 
In this note we discuss possible evidence for an enrichment by field SNe 
from the ratio of well localized gamma ray bursts (GRBs) with and without 
host galaxies, as well as from GRBs coincident with SNe but without a host 
galaxy. A direct proof of our contention that the enrichment of iron in 
the IGM is partly due to `hostless' field supernovae (SNe not within galaxies) 
would require searches with IR telescopes --such as the Spitzer Space 
Telescope-- and measurements of their rate per unit volume as function of 
$z$.

\section{Iron production by ordinary Supernovae}

In the simplest `bottom-up' structure-formation models, stars form first. 
Subsequently, `assisted' by dark matter,
they form galaxies.  Finally, galaxies `separate from the universal
expansion' and become galaxy clusters\footnote{In this picture, the
ratio of baryonic to dark matter mass in galaxy clusters is the same as 
that in the whole universe, and the total cluster's masses are 
proportional to their light, as 
observed.}. The progenitors of SNe are massive stars of 
short lifetime. Thus, the SN rate should follow the star 
formation rate. The observed rate, 
${\rm SFR}(z)$, 
is well represented (e.g., Perez-Gonzalez et al.~2005; 
Schiminovich et al.~2005) by 
${\rm SFR}(z)\!=\!{\rm SFR}(0)\,(1+z)^4$ 
for $z\leq 1$, and ${\rm SFR}(z)\!\approx\! {\rm SFR}(z=1)$ for $1\leq z\leq 5$, 
see Fig.~1.

Galaxy clusters have been formed quite recently. Since star formation took 
place mainly during the pre-cluster stage, the IGM and ICM are expected to 
have similar iron abundances, produced during the whole history of star 
formation. Indeed, observations indicate that most of the cluster metals 
were produced at $z>1$ (e.g., Mushotzky \& Loewenstein~1997; Tozzi et 
al.~2003). Yet, SN explosions, the main known sources of iron in the 
universe, take place (Gal-Yam \& Maoz~2003) both in the ICM and in the 
galaxies of the cluster (mainly SNe of type Ia in elliptical galaxies, and 
in the CD galaxy dominating the centers of the cooling-core clusters and 
producing the observed central enrichment).
 
Supernovae of type Ia are believed to be  
thermonuclear explosions of white dwarfs whose mass exceeds the 
Chandrasekhar limit due to mass accretion or  merger in close binaries .
Their Fe mass yield is $\approx 
0.7M_\odot$ per SN, and their local rate is $(0.37\pm 0.11)\,h^2\, {\rm SNU}$,
where $h$ is the Hubble constant $H_0$ in units of
$100\, {\rm km\, s^{-1}\, Mpc^{-1}}$ and SNU is the number of SNe
per century and per a luminosity of
$10^{10}\, L_{B,\odot}$ (Capellaro et al.~1999).
The progenitors of white dwarfs of a mass near the Chandrasekhar 
limit are
probably short lived (relatively to the Hubble expansion rate)
massive stars with a mass slightly less than $\sim 8\, M_\odot\, ,$ 
whereas 
the progenitors of core-collapse SNe are believed to be 
stars more massive than $\sim 8\,M_\odot\,.$
The observed Fe mass yield of core-collapse SNe (supernovae of types Ib/c 
and II) is in the range  0.0016 to $ 0.26\, M_\odot$ (Hamuy~2003)
with a mean value  $\approx (0.05\pm 
0.03)\,M_\odot$ (Elmhamdi, Chugai \& Danziger~2003).
Their local rate is $(0.85\pm 0.35)\,h^2$ SNU. 
Thus, for $h=0.73$,
the  rate of iron production per unit volume in the local 
universe, which follows from the local 
luminosity density of the universe in the B band (Cross et al.~2001), 
$j_{_B}\approx (2.49 \pm 0.20)\, 10^8\,h\, L_{B,\odot}\, {\rm Mpc}^{-3}$,   
is:
\begin{equation}
\dot{M}_{Fe}\approx (0.30 \pm 0.10)\, h^2\,{\rm SNU} \,M_\odot\, j_{_B}\approx 
(0.29\pm 0.10)\times 10^{-4}\, M_\odot\, 
\rm Mpc^{-3}\, y^{-1}\,.
\label{RMFe}
\end{equation}
For a SN  rate proportional to the star formation rate,
Eq.~(\ref{RMFe}) yields an iron number density:
\begin{equation}
n_{_{Fe}}\approx {(0.28\pm 0.10)\times 10^{-4}\, M_\odot \over 56\, m_p\, 
            {\rm Mpc^3\,y}}\, \int {{\rm SFR}(z)\over {\rm SFR}(0)}\,{dt\over dz}\,dz\,.
\label{nFe}
\end{equation}
In a standard flat universe
\begin{equation}
          {dt\over dz}={1\over H_0\, (1+z)\, 
           \sqrt{(1+z)^3\, \Omega_M+\Omega_\Lambda}}
\label{dtdz} 
\end{equation} 
where $H_0=(73\pm 2)\, {\rm km\, s^{-1}\, Mpc^{-1}}$, and 
$\Omega_M=0.24\pm 0.02 $ and $\Omega_\Lambda=0.76\pm0.02$ 
are, respectively, the matter density and the density of dark energy in 
critical units, $\rho_c=3\,H_0^2/8\, \pi\, G \approx (1.00 
\pm 0.05) \times 10^{-29}\, 
{\rm g\, cm^{-3}}$, as inferred 
from the measurements of the anisotropy of the microwave background
radiation (Spergel et al.~2006) with the 
Wilkinson Microwave Anisotropy Probe. With these parameters, 
Eq.~(\ref{nFe}) yields a mean density of iron nuclei in the present 
universe, $n_{_{Fe}}\approx (2.8\pm 1.0) \times 10^{-12}\,{\rm cm}^{-3}$,
where the error is dominated by the uncertainty in the rate
of SNIa. The current 
mean cosmic baryon density in critical units as inferred from the sky 
distribution of the cosmic microwave background radiation 
(Spergel et al.~2006) is $\Omega_b=0.042 \pm 0.002$,
resulting in a mean cosmic baryon 
density of $n_b= (2.52 \pm 0.2) \times 10^{-7}{\rm cm}^{-3}$, consistent 
with its inferred value from Big Bang Nucleosynthesis. 
The hydrogen mass fraction produced by Big 
Bang nucleosynthesis is, $\approx 76\%\pm 1\%$ (e.g., Peimbert et 
al.~2007), and has not changed much by stellar evolution. 
Approximately 1/6 of the baryons in galaxy clusters reside  within 
galaxies, while 5/6 are in the ICM (e.g., Ettori~2003). Hence, the 
mean ratio of iron nuclei to hydrogen nuclei in the IGM is:
\begin{equation}
    [Fe/H]_{IGM} \approx (1.22\pm 0.3) \times 10^{-5}\,.
\label{RFeH}
\end{equation}
The relative abundance of iron to hydrogen in the Sun (Grevesse \& Sauval~1998)
is $[F/H]_\odot\approx (3.16 \pm 0.16)\times 10^{-5}\,,$ and the 
expected mean iron abundance in the IGM from SNe,  given by 
Eq.~(\ref{RFeH}), satisfies $[Fe/H]_{IGM}\approx (0.38\pm 0.14)\times 
[F/H]_\odot$,  in  agreement with  observations.

\section{IGM Metallicity From Galactic Injection}
In the local universe, the total mass of the gas in spiral and 
elliptical galaxies is much smaller than the total mass in stars.
Approximately 1/6 of the baryons in 
galaxy clusters reside in stars  within
galaxies, while 5/6 are in the ICM (e.g., Ettori~2003).
The simplest  interpretation of this observation is is that  
most of the gas in galaxies   
was swept into the IGM by strong winds and jets produced in star
formation and stellar evolution. Consequently, these winds and jets,  
have transported  into the IGM the the metals that were injected into the 
ISM by SN explosions.

\section{IGM Metallicity From Hostless SNe}

There is mounting evidence that most, perhaps all long duration GRBs are 
associated with 
core-collapse supernova explosions (for a review see,
e.g., Dar~2004.)
Evidence for such an association was already visible in the 
first discovered optical afterglow of a GRB, i.e.~that of GRB 970228, but 
became convincing only after the measurement of its red-shift (Dar~1999; 
Reichart~1999; Galama et al.~2000). The first clear 
evidence for a 
GRB-SN association came from the discovery of SN1998bw in the error circle of 
GRB980425 by Galama et al.~1998). It was not widely accepted and was 
argued to be either a chance coincidence or  an association 
between a rare type of GRB and a rare type of SN. But shortly after, 
evidence for a SN contribution to the late optical afterglow of 
an ordinary GRB (980326)
was discovered by Bloom et al.~1999). 
The late bump in its optical afterglow, if it was produced by a bright SN 
akin to SN1998bw, indicated that the red-shift of the GRB/SN was less than 
1. However, deep searches with HST for a host galaxy failed to detect it 
down to $V=29.25\pm 0.25$ (Fruchter et al.~2001) --
within one pixel of the estimated position, there was $\sim 4.5\, \sigma$
evidence of a small 
source of this magnitude. A 
galaxy at $z\sim 1$ with this observed magnitude is 7 magnitudes below 
$L^*$, the knee of the galaxy luminosity function at that red-shift. 
GRB 980326 was the first GRB to provide evidence for production 
of GRBs by  SNe perhaps without a host galaxy (field SNe).

A nearly 1:1 correspondence between long-duration GRBs and core-collapse
SNe is compatible with the observations (e.g., A.~Dar 2004).
But the afterglows of two recent long GRBs, 060614 and 060505
(see, however, Ofek et al.~2007), which 
were located in the outskirts of nearby galaxies, did not show any evidence 
for an associated SN (Gal-Yam et. al.~2006; Fynbo et al.~2006).
Although it is 
conceivable that these GRBs belong to a new class of GRBs (Gal-Yam et 
al.~2006), e.g., GRBs associated with `failed' SNe, as conjectured 
long ago by Woosley~1993, it was pointed out by Schaefer and 
Xiao~(2006) and Dado et al.~2006) that GRB 060614 looks like an ordinary GRB at 
a much higher red-shift, 
$z\sim 1.9,$ as suggested by various red-shift indicators of ordinary GRBs.
The proximity of the sky position of these two GRBs to nearby 
galaxies could have been a chance coincidence and not a physical 
association. An underlying SN at red-shift $z\sim 1.9$ is too dim to be 
detected in the afterglow of an ordinary GRB at a red-shift $z\sim 1.9$. 
Discussing GRB 060614,
Gal-Yam et al.~2006) responded that deep searches with the HST failed to 
detect a host galaxy at $z\sim 1.9$. However, if this GRB was an 
ordinary one, produced by a field SN at that red-shift, there
would be no galaxy to be observed by the HST at its sky position.

The current ratio between the number of baryons in galaxies and in the ICM 
of galaxy clusters, inferred from X-ray and optical observations, is 
$\sim$ 5:1, e.g., Ettori~2003). The metallicity of the ICM 
of galaxy clusters is $\sim 3$ times less than in their galaxies, If all 
this metallicity was produced by SNe outside galaxies, we would expect the 
ratio of SNe which took place in the IGM to those which took place in 
galaxies to be roughly 2:1. Such a ratio is not ruled out by the observed 
ratio of GRBs with and without a detected host galaxy: out of the 97 SWIFT 
GRBs which were well localized todate by their optical and/or radio 
afterglows, nearly 2/3 have no detected host galaxy although in many cases 
searches were not deep enough to rule out a host galaxy to a strong limit.

\section{Conclusions}
 
The simple calculation presented in this note demonstrates that no early 
iron enrichment of the intracluster gas by Population III stars is 
required by the observed ICM metallicity in the local universe. To reach 
this conclusion, we have made the very reasonable assumption that the 
rates of thermonuclear and core-collapse SNe, in the gas which ended in 
the ICM of clusters, were proportional to the cosmic star-formation rate. 
Our results agree with the observed metallicities in the ISM in galaxies, 
in the ICM, and in the IGM. The metals in the IGM within and without 
galaxy clusters could have been transported there with most of the matter 
in the ISM by Galactic winds and jets during stellar evolution and 
partially produced there by intergalactic SNe.  Combined with the observed 
levels of the associations between (mainly long-duration) GRBs and SNe 
(nearly 1:1), as well as between GRBs and host galaxies, our results imply 
that up to 2/3 of the long-duration GRBs could have been produced by such 
{\it field} SNe. This is consistent with the current data on long-duration 
GRBs, but deeper searches for host galaxies of well-localized long GRBs 
would further test it.  A few observed cases of GRBs or X-ray flashes 
associated with SNe, but with no detected host galaxy, support our 
conclusion. A conclusive test whether a considerable iron enrichment of 
the intergalactic medium was by field SNe would require measurements of 
their rate per unit volume and red-shift with IR telescopes, such as the 
Spitzer Space Telescope. Such searches may settle the questions of the 
cosmic distribution of SNe and of the need to assume the existence of a 
first generation of Population III stars.

{\bf Acknowledgment:} We thank S. Covino and A. Gal-Yam for useful 
comments. This research was supported in part by the Asher Space Research 
Fund at the Technion.

\newpage
\begin{figure}
   \begin{center}
   \epsscale{1.0}{1.0}          
   \plotone{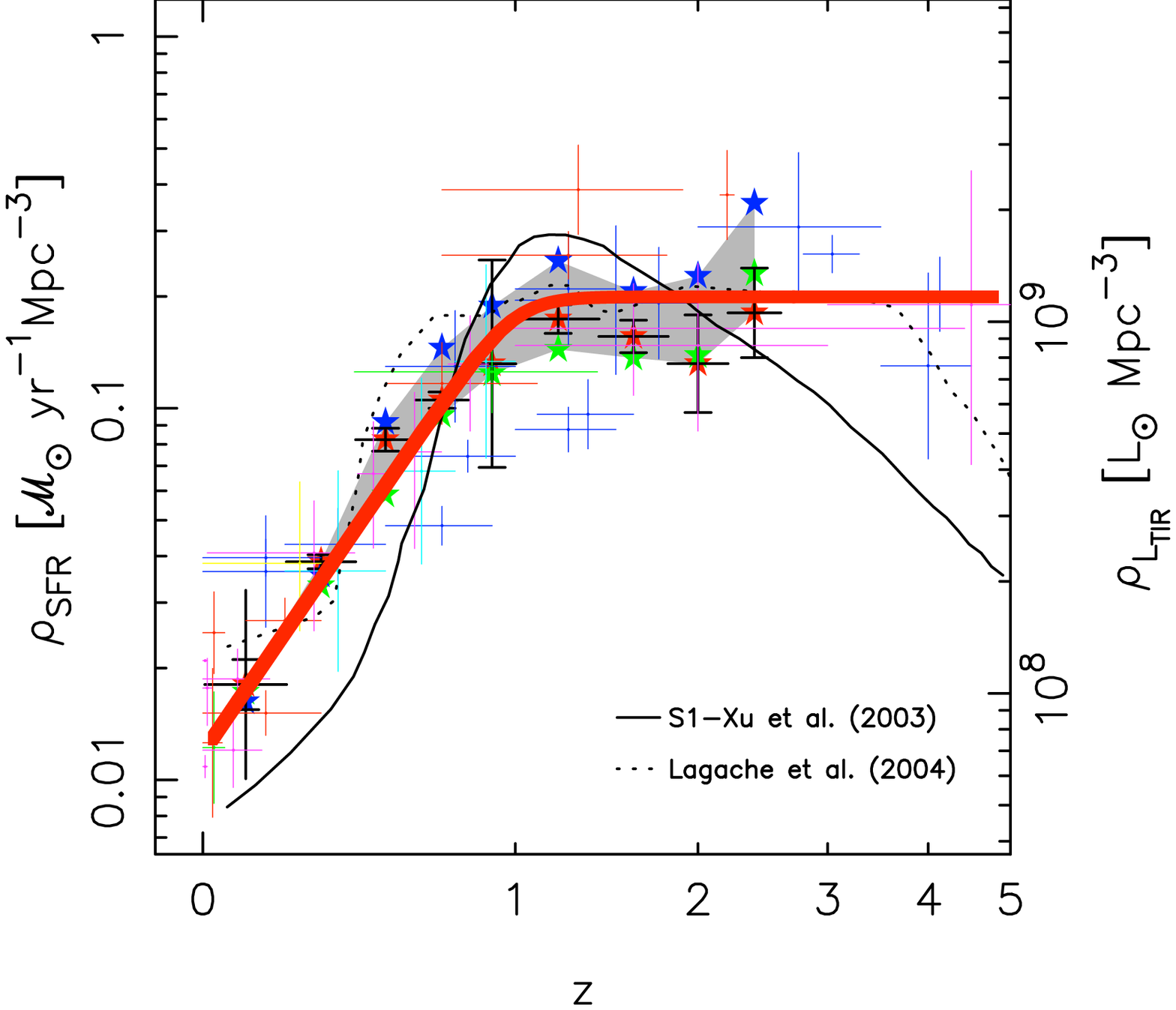}
   \caption{The star formation rate
    as function of red-shift,  compiled by Perez-Gonzalez et al.~(2005).
The colored points (shown with error bars) are extracted from different 
sources in the literature, normalized to the same standard cosmology.
Red symbols are estimations based on ${\rm H\alpha}$ or
 ${\rm H\beta}$ measurements. Green symbols stand for [OII]$\lambda$3737 
estimations. UV-based data points are plotted in blue.
Cyan estimations are based on mid-infrared data.   
The yellow point is based on X-ray data.
The shaded area delimits the zone between two extreme SFR density 
estimations for each redshift.
The horizontal bars show the range of redshifts used in each bin  
The curves show two typical models: one 
with a decay from z=1, and another with a 
constant SFR density at high redshift.
The thick (red) line is used in our calculations.}
   \label{FigSFR}
   \end{center}
\end{figure}


\begin{thebibliography}{99}

\bibitem[1992]{Arnaud1992} 
Arnaud, M., et al., 1992, A\&A, 254, 49
\bibitem[2007]{Balestra2007}
Balestra, I., et al. 2007, astro-ph/0703261
\bibitem[2005]{Barkats2005} 
Barkats, D., et al. 2005, ApJ, 619, L127
\bibitem[1999]{Bloom1999} 
Bloom, J. S., et al. 1999, Nature, 401, 453 
\bibitem[1998]{Brighenti1998} 
Brighenti, F. \&  Mathews, W. G. 1998, ApJ, 514, 542
\bibitem[1999]{Capellaro1999} 
Capellaro, E., Evans. R. \& Turatto, M. 1999 A\&A 351, 459 
\bibitem[2006]{Dado2006} 
Dado, S., Dar, A., \& De R\'ujula, A. 2006, astro-ph/0611161
\bibitem[1999]{Dar1999} 
Dar, A. 1999, GCN No 346
\bibitem[2004]{Dar2004} 
Dar, A. 2004, astro-ph/0405386
\bibitem[2001]{Cross2001} 
Cross, N., et al. 2001, MNRAS, 324, 825
\bibitem[1991]{Edge1991} 
Edge, A. C. \& Stewart, G. C. 1991, MNRAS, 252, 414 
\bibitem[2003]{Elmhamdi2003} 
Elmhamdi, A., Chugai, N. N. \& Danziger, I. J. 2003, A\&A, 359, 876 
\bibitem[2003]{Ettori2003} 
Ettori, S., 2003, MNRAS, 344, L13
\bibitem[2003]{Fosbury2003} 
Fosbury, R. A. E., et al. 2003, ApJ, 596, 797 
\bibitem[2001]{Fruchter2001}
Fruchter, A., et al., 2001, GCN No. 1029
\bibitem[2004]{Fynbo2004} 
Fynbo, J. P. et al., 2004,  ApJ, 609, 692
\bibitem[2006]{Fynbo2006} 
Fynbo, J. P. et al., 2006,  Nature, 444, 1047
\bibitem[1998]{Galama1998}
Galama, T. J., et al.,  1998, Natur, 395, 670 
\bibitem[2000]{Galama2000} 
Galama, T.J., et al., 2000, ApJ, 536, 185 
\bibitem[2003]{Gal-Yam2003}
Gal-Yam, A., et al. 2003, AJ, 125, 1087
\bibitem[2006]{Gal-Yam2006} 
Gal-Yam, A., et al. 2006, Nature, 444, 1053 
\bibitem[1998]{Grevesse1998} 
Grevesse, N. \& Sauval, A. J. 1998, Space Sci. Rev. 85, 161 
\bibitem[2003]{Hamuy2003} 
Hamuy, M., 2003, ApJ, 582, 905 
\bibitem[2003]{Kawai2003} 
Kawai, N., et al., 2003, GCN No. 2412
\bibitem[2001]{Loeb2001} 
Loeb, A. \& Barkana, R. 2001, ARAA, 39, 19
\bibitem[2005]{Mannucci2005}
Mannucci, F., et al. 2005, A\&A, 433, 807
\bibitem[2007]{Mannucci2007}
Mannucci, F., Della Valle, M. \& Panagia, N. 2007, astro-ph/0702355
\bibitem[2004]{Maoz2004}
Maoz, D. \& Gal-Yam, A.,  2004, MNRAS 347, 951 
\bibitem[1997]{Mushotzky1997} 
Mushotzky, R.F., Loewenstein, M. 1997, ApJ, 481, L63
\bibitem[2007]{Ofek2007}
Ofek, E. O., et al. 2007, astro-ph/0703192
\bibitem[2005]{Perez2005} 
Perez-Gonzalez P. G.,  et al. 2005, ApJ, 630, 82
\bibitem[2007]{Peimbert2007} 
Peimbert, M., Luridiana, V. \& Peimbert, A. 2007, astro-ph/0701580 
\bibitem[2003]{Scannapieco2003} 
Scannapieco, E.,  Schneider, R. \&  Ferrara, A. 2003, ApJ, 589, 35 
\bibitem[1999]{Reichart1999} 
Reichart, D. E. 1999, ApJ, 521, L111
\bibitem[1993]{Renzini1993} 
Renzini, A., et al., 1993, ApJ. 488, 35 
\bibitem[2006]{Schaefer2006} 
Schaefer, B. E. \& Xiao, L. 2006, astro-ph/0608441
\bibitem[2005]{Schiminovich2005}
Schiminovich, D. et al. 2005, ApJ, 619, L47
\bibitem[2006]{Spergel2006} 
Spergel, D. N., et al. 2006, ApJS in press (astro-ph/0603449) 
\bibitem[2003]{Tozzi2003} 
Tozzi, P., et al. 2003, ApJ, 593, 2003
\bibitem[1993]{Woosley1993} 
Woosley, S. E., 1993, ApJ, 405, 73
\end{thebibliography}
\end{document}